\documentclass[12pt, draftclsnofoot, onecolumn]{IEEEtran}
\usepackage{balance}
\usepackage{cite}
\usepackage{graphicx}
\usepackage{psfrag}
\usepackage{subfigure}
\usepackage{url}
\usepackage{stfloats}
\usepackage{amsmath}
\usepackage{array}
\usepackage{amssymb}
\usepackage{setspace}
\usepackage{algorithmicx}
\usepackage[ruled]{algorithm}
\usepackage{algpseudocode}
\usepackage{algpascal}
\usepackage{algc}
\usepackage{color}

\begin{document}

\title{Capacity Enhanced Cooperative D2D Systems over Rayleigh Fading Channels with NOMA}
\author{Wei Duan, Jinjuan Ju, Qiang Sun, Yancheng Ji, Zhiliang Wang,\\ Jeaho Choi, and Guoan
Zhang$^{\dag}$
\thanks{W. Duan, J. Ju, Q. Sun, Y. Ji, Z. Wang and G. Zhang are with School of Electronics and Information,
Nantong University, Nantong 226019, China. (e-mail: \{sinder,
sunqiang, jiyancheng, wangzl, gzhang\}@ntu.edu.cn)}
\thanks{J. Choi is Division of Electronics Engineering,
Chonbuk National University, Korea. (e-mail: wave@jbnu.ac.kr)}
\thanks{J. Ju is also with School of Electronics and Information,
Nantong University, Nantong 226019, China. (e-mail:
Janicejjj@163.com}
\thanks{$^\dag$Corresponding email: gzhang@ntu.edu.cn.}
} \maketitle

\begin{abstract}

This paper considers the cooperative device-to-device (D2D)
systems with non-orthogonal multiple access (NOMA). We assume that
the base station (BS) can communicate simultaneously with all
users to satisfy the full information transmission. In order to
characterize the impact of the weak channel and different decoding
schemes, two kinds of decoding strategies are proposed:
\emph{single signal decoding scheme} and \emph{MRC decoding
scheme}, respectively. For the \emph{single signal decoding
scheme}, the users immediately decode the received signals after
receptions from the BS. Meanwhile, for the \emph{MRC decoding
scheme}, instead of decoding, the users will keep the receptions
in reserve until the corresponding phase comes and the users
jointly decode the received signals by employing maximum ratio
combining (MRC). Considering Rayleigh fading channels, the ergodic
sum-rate (SR), outage probability and outage capacity of the
proposed D2D-NOMA system are analyzed. Moreover, approximate
expressions for the ergodic SR are also provided with a negligible
performance loss. Numerical results demonstrate that the ergodic
SR and outage probability of the proposed D2D-NOMA scheme
overwhelm that of the conventional NOMA schemes. Furthermore, it
is also revealed that the system performance including the ergodic
SR and outage probability are limited by the poor channel
condition for both the \emph{single signal decoding scheme} and
conventional NOMA schemes.

\end{abstract}
\textbf{Index Terms:}  Non-orthogonal multiple access (NOMA),
device-to-device (D2D), ergodic sum-rate, decode-and-forward (DF),
outage probability, outage capacity.

\section{Introduction}

For the next generation networks, with the massive connectivity of
cellular internet-of-things (IoT) \cite{m1}, non-orthogonal
multiple access (NOMA) has been envisioned as a promising
technique to realize the spectral efficient massive access
\cite{m2,m3}. Unlike conventional orthogonal multiple access (OMA)
schemes such as frequency division multiple access (FDMA), time
division multiple access (TDMA), and code division multiple access
(CDMA), etc., NOMA allows multiple users to simultaneously
transmit signals in a superposition way by using the same
orthogonal resources (i.e., time/frequencey/code) but different
power levels \cite{m4}. The key idea of NOMA is to explore the
power domain for realizing multiple access, where the users with
strong channels are normally limited by the bandwidth, in the
meanwhile that the users with weak channels are limited by the
noise. In summary, the concept of NOMA is to superpose multiple
users over different power levels at the transmitter, and employ
the successive interference cancellation (SIC) to detect the
signals at the receiver. More specifically, an overview of the
NOMA researches and future trends has been studied in \cite{ii1},
which shows that massive connectivity can be realistically
achieved with NOMA and the delay can be also reduced since the
users are no longer forced to wait until an orthogonal resource
block becomes available. In \cite{ii2}, the authors
comprehensively surveyed the recent progress of NOMA in 5G
systems, including the state-of-the-art capacity analysis, power
allocation strategies, user fairness, and user-pairing schemes in
NOMA. Furthermore, an overview of the resource allocation
algorithms for downlink NOMA in a categorized fashion is studied
in \cite{ii3}, which provides the basis for further research
directions on the resource allocation for NOMA systems, such as
the joint optimization of power allocation (PA), lower complexity
resource allocation, and security-aware resource allocation.

In the recent researches, for the excellent compatibility of the
NOMA with other communication technologies, NOMA has been
integrated in various communication systems. Since massive
multiple-input multiple-output (MIMO) \cite{y1,y2,y3} makes a
clean break with current practice through the use of a very large
number of service antennas, in \cite{y6}, the authors proposed a
massive-MIMO-NOMA transmission scheme. Particularly, in \cite{b3,
b4, h1}, the authors studied the secrecy issue of NOMA for single-
and multiple-antenna scenarios, where the security performance of
the NOMA networks can be significantly improved compared with the
convention OMA one. In \cite{e1}, the resource allocation for the
downlink MIMO-based NOMA (MIMO-NOMA) system is studied, in which,
the proposed system achieves near optimal sum-rate (SR)
performance with a high-complexity beamforming scheme by
considering both perfect and imperfect channel state information
(CSI). Since radio-frequency (RF) signals have the ability to
carry both information and energy, simultaneous wireless
information and power transfer (SWIPT) has been attracted
increasing attention in the communication research community
\cite{gg1}. Motivated by the advantages of NOMA and SWIPT, the
application of SWIPT to NOMA networks are investigated in
\cite{e2, c2}, where the user with a strong channel condition
serves as an energy-harvesting (EH) relay to complete the power
splitting scheme. In addition, the antenna and relay selection
problems for NOMA systems have been studied in \cite{w1, w2},
which demonstrate that the proposed selection schemes yield a
significant performance gain over the OMA. More specifically, the
works \cite{w3, w4} introduced NOMA into short packet
communication to achieve low latency and a much higher effective
throughput, as well as reduce the latency in short-packet
communications for achieving the same effective throughput
compared with the OMA. It is worth noting that, the fairness is
also an important issue for NOMA \cite{g2}, since there is a
tradeoff between the total throughput and user fairness. Moreover,
in \cite{ll1}, since cooperative NOMA systems inherit advantages
of the NOMA protocol and cooperative relaying, the cooperative
NOMA systems in MIMO channels are considered, which outperforms
the conventional NOMA scheme in terms of the achievable rate.
Specifically, the imperfect CSI scheme with NOMA is also studied
in \cite{ll2}, where the robust scheme is developed based on the
worst-case performance optimization framework.

Since the relaying transmission significantly increases the system
capacity, the cooperative relaying networks (CRNs) have garnered
considerable interest. Recently, the CRN-NOMA systems have been
proposed in the literatures \cite{ttt1,f1,f2,f3,a4,a11,a12,a13,
a14}. The authors in \cite{f1} investigated a CRN over Rician
fading channels by deriving the exact analytical expressions of
the achievable rates. It is noted that in \cite{f2}, the outage
performance of CRN-NOMA with an amplify-and-forward (AF) relay is
studied, which reveals the proposed CRN-NOMA achieves the same
diversity order and a superior coding gain compared with the
cooperative OMA scheme. Considering an orthogonal frequency
division multiplexing (OFDM) AF relaying system allocated the
spectrum and power resources, the resource allocation problem for
a single-cell CRN-NOMA is studied in \cite{f3}, which results in a
maximum average SR. In particular, adopting maximal ratio
combining (MRC), NOMA for multiple-antenna relaying networks are
designed in \cite{a4} which shows the correctness of the
theoretical analysis and the superiority of NOMA. Meanwhile, a
coordinated direct and relay transmission (CDRT) with the
decode-and-forward (DF) protocol in NOMA system is investigated in
\cite{a11}, in which, it is also showed that the main challenges
of the non-orthogonal CDRT can be solved by using the inherent
property of NOMA. It is worth noting that, the achievable ergodic
rate is restricted by the weak channel for its lower channel gain.
To solve this problem, the authors in \cite{a12} designed a novel
MRC receiver which provides an excellent performance gain in terms
of the ergodic SR and outage probability. Unfortunately, the MRC
in \cite{a12} requires the lower power signals to be decoded first
that does not strictly follow the NOMA decoding principle. With
these observations, \cite{a13} proposed a two-stage power
allocation scheme, where a novel type of the construction for the
superposition coded signals at the relay is designed to avoid the
problem described above. Moreover, a device-to-device (D2D) aided
NOMA is proposed in \cite{a14}, in which, the spectral efficiency
is significantly improved for both the D2D aided cooperative
communication and D2D communication systems.

Actually, in \cite{a14}, the signals forwarded by the relay node
do not include the full information (i.e., the signal $x_1$ is not
forwarded during the second phase), in the meanwhile that there is
no direct link between the base station (BS) and user 3 (UE3). On
the other hand, it is possible to see that, with the similar
transmission and decoding strategies, straightforwardly including
one direct link between the BS and UE3 to satisfy the full
information transmitting requirements, the performance of the
inspired systems will be worse than that the one in \cite{a14} due
to the poor channel limitation (the path loss from the BS to users
is normally worse than that of the one from the relay to users).
In this paper, we will first characterize the impact of the weak
channel and different decoding schemes, and then provide the
mentality to study the D2D-NOMA systems, meanwhile the applicable
in the future cooperative NOMA networks. These motivate us to
investigate a capacity enhanced transmission scheme for the
proposed D2D-NOMA systems, which considering the superposition
coded signals are transmitted from both of the BS and relay. More
details of implementations and essential contributions of this
paper are summarized as follows:
\begin{itemize}
\item Considering all the users are within the transmission scope
of the BS, to further improve the spectrum efficiency, the direct
links between the BS and users are necessary to be included. For
the proposed D2D-NOMA system, unlike existing works, a more
practical D2D scenario is proposed and investigated, wherein the
BS can communicate simultaneously with all the users to satisfy
the full information transmitting requirements. In particular,
both the BS and relay are allowed to transmit superposition coded
signals. \item Without loss of the generality, to characterize the
impact of the weak channel condition and different decoding
schemes, two decoding strategies are proposed which are termed as
\emph{single signal decoding scheme} and \emph{MRC decoding
scheme}. For the \emph{single signal decoding scheme}, all the
users, i.e., the BS and relay, will immediately decode the
received signals after receptions from the BS. Unlike the
\emph{single signal decoding scheme}, for the \emph{MRC decoding
scheme}, instead of decoding, the users will keep the receptions
in reserve until the corresponding phase comes. \item Asymptotic
closed-form expressions for the ergodic SR, outage probability and
outage capacity of the proposed D2D-NOMA systems are derived with
a negligible performance loss for high transmit SNR scenarios,
respectively. The correctness of the analytical results is
corroborated by simulations. Moreover, we also discuss the ergodic
SR judgement of these two decoding schemes through the asymptotic
closed-form expressions. \item By means of the numerical results,
both analytically and numerically, we compare the proposed
D2D-NOMA schemes with the one in \cite{a14} in terms of the
ergodic SR and outage probability. It is demonstrated that, the
proposed \emph{MRC deocing scheme} outperforms that of the one in
\cite{a14} and \emph{single signal decoding scheme} significantly.
In addition, it is also shown that the system performance will be
limited by the poor channel condition for the \emph{single signal
decoding scheme} and conventional NOMA scheme, but not for the
proposed \emph{MRC decoding scheme}.
\end{itemize}

The remainder of this paper is organised as follows. In Section
II, the system mode for the proposed D2D-NOMA system is
introduced. In Section III, the analytical expressions of the
achievable ergodic SR for the proposed two decoding strategies are
derived. Then the outage probability and outage capacity for
D2D-NOMA are investigated in Section IV. Analytical results and
numerical simulations are presented in Section V. Finally, Section
V concludes this paper.

\emph{Notations:} $\mathcal{CN}\left(\cdot\right)$ represents a
complex Gaussian distribution. $\mathrm{Ei}\left(\cdot\right)$,
$\mathrm{Ec}$ and $\mathrm{E}\left[\cdot\right]$ denote the
exponential integral function, Euler constant and expectation,
respectively. $\Pr\left\{A|B\right\}$ denotes the conditional
probability of the event $A$ on event $B$.
$\left|\cdot\right|^{2}$ stands for the norm square of a scalar.

\section{System model and proposed scheme}

\begin{figure}
\begin{center}
\includegraphics [width=150mm]{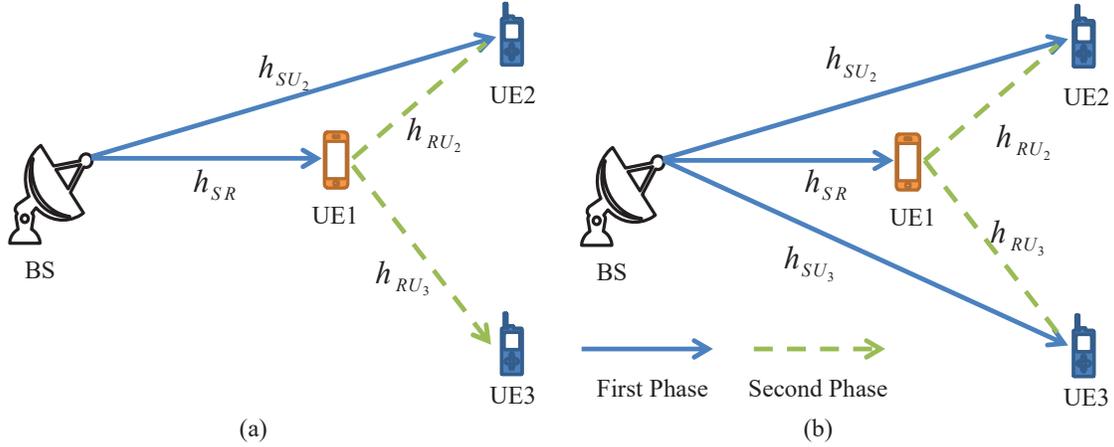}
\caption{\label{9} (a) The convention D2D-NOMA system. (b) The
proposed capacity enhanced cooperative D2D system with NOMA.}
\end{center}
\end{figure}

By employing decode-and-forward (DF) schemes, a simple cooperative
D2D relaying system consisting of one BS, one relay (UE1) and two
users (UE2 and UE3) is considered as shown in Fig. 1, in which all
nodes operate in a half-duplex mode. It is also assumed that all
the nodes in the system are equipped with a single antenna,
meanwhile that all receiving nodes acquire the perfect CSI. The
independent Rayleigh fading channel coefficients from the BS to
users $i\in\left\{1,2,3\right\}$, and from the relay to UE2 and
UE3 are denoted as $h_{BU_{i}}$, $h_{RU_{2}}$ and $h_{RU_{3}}$,
where the average powers are given as $\alpha_{BU_{i}}$,
$\alpha_{RU_{1}}$ and $\alpha_{RU_{2}}$, respectively. Without
loss of generality, it is also assumed that
$\alpha_{BU_{i}}<\min\{\alpha_{RU_{1}},\alpha_{RU_{2}}\}$, since
that the path loss from the BS to users is normal worse than that
of the one from the relay to users.

In our proposed scheme, two-phase transmission is considered. With
the observation of the NOMA principle, the power allocation factor
$a_{i}$, for $i\in\left\{1,2\right\}$, follows the conditions of
$a_1>a_2$ and $a_{1}+a_{2}=1$, which are related to the quality of
the channel coefficients. During the first phase, a superposed
signal
\begin{eqnarray}
\sqrt{a_{1}P_{t}}x_{1}+\sqrt{a_{2}P_{t}}x_{2}
\end{eqnarray}
is transmitted from the BS to user $i$, where $x_{i}$ denotes the
broadcasted symbol at the BS and $P_{t}$ means the total transmit
power. Therefore, the received signal at the user $i$ can be given
as
\begin{eqnarray}\label{1}
y_{U_{i}}&=&h_{SU_{i}}\left(\sqrt{a_{1}P_{t}}x_{1}+\sqrt{a_{2}P_{t}}x_{2}\right)+n_{U_{i}},
\end{eqnarray}
where $n_{U_{i}}\sim{\mathcal{CN}(0,\sigma^{2})}$ stands for the
additive white Gaussian noise (AWGN) with zero mean and variance
$\sigma_{U_{i}}^{2}$.

In order to successfully and simultaneously decode the symbols
$x_{1}$ and $x_{2}$ at UE1, the SIC technique is utilized.
Following this manner, the received SNRs for $x_{1}$ and $x_{2}$
at UE1 can be respectively obtained as
\begin{eqnarray}\label{2}
\gamma_{U_{1}}^{\left(x_{1}\right)}=\frac{\left|h_{SU_{1}}\right|^{2}a_{1}\rho}{\left|h_{SU_{1}}\right|^{2}a_{2}\rho+1}
~,~
\gamma_{U_{1}}^{\left(x_{2}\right)}=\left|h_{SU_{1}}\right|^{2}a_{2}\rho,
\end{eqnarray}
with $\rho=\frac{P_{t}}{\sigma^{2}_{R_{i}}}$ as the transmit SNR.

During the second phase, for UE1, to improve the spectral
efficiency as well as offload the traffic, our proposed capacity
enhanced D2D-NOMA systems allow the relay to forward its own
signal to the UE2 and UE3. With this observation, the relay node
forwards a new superposition coded to the destination:
\begin{eqnarray}\label{h1}
\sqrt{b_1}x_{2}+\sqrt{b_2}x_{r}
\end{eqnarray}
where $b_{i}$, for $i\in\left\{1,2\right\}$, is the new power
allocation coefficient with $b_1+b_2=1$. It is worth noting that,
since the signal $x_2$ is with lower transmit power during the
first phase, which leads to a worse spectral efficiency at the UE2
and UE3. Therefore, during the second phase, the signal $x_2$ is
reasonable to be retransmitted to solve this problem.
Correspondingly, the received signals at the UE2 and UE3 can be
expressed as
\begin{eqnarray}\label{3}
r_{U_{2}}&=&h_{RU_{2}}\left(\sqrt{a_{1}P_{t}}x_{2}+\sqrt{a_{2}P_{t}}x_{r}\right)+n^{(2)}_{U_{2}},\notag\\
r_{U_{3}}&=&h_{RU_{3}}\left(\sqrt{a_{1}P_{t}}x_{2}+\sqrt{a_{2}P_{t}}x_{r}\right)+n^{(2)}_{U_{3}},
\end{eqnarray}
where $n^{(2)}_{U_{2}}$ and $n^{(2)}_{U_{3}}$ are the AWGN with
zero mean and variance $\sigma_{U_{2}}^{2}$ and
$\sigma_{U_{3}}^{2}$, respectively.

Without loss of generality, in this paper, two kinds of decoding
strategies will be studied: one is similar to \cite{a14}, the
other is the MRC one, which will be introduced in the following
subsections.

\subsection{Single Signal Decoding Scheme}

During the first phase, the UE2 and UE3 immediately decode the
receptions, i.e., $x_1$, with the corresponding received SNR as
\begin{eqnarray}\label{m1}
\gamma_{U_l}^{\left(S,x_{1}\right)}=\frac{a_1\left|h_{SU_l}\right|^{2}\rho}{a_2\left|h_{SU_l}\right|^{2}\rho+1}.
\end{eqnarray}
At the second phase, the UE2 and UE3 will respectively decode the
received signals $x_2$ and $x_r$. By this way, the received SNRs
for $x_2$ and $x_r$ are given as
\begin{eqnarray}\label{m2}
\gamma_{U_l}^{\left(S,x_{2}\right)}=\frac{b_1\left|h_{RU_l}\right|^{2}\rho}{b_2\left|h_{RU_l}\right|^{2}\rho+1},
\end{eqnarray}
and
\begin{eqnarray}\label{m3}
\gamma_{U_l}^{\left(S,x_{r}\right)}=b_2\left|h_{RU_l}\right|^{2}\rho.
\end{eqnarray}
In this manner, the achievable SR can be obtained from
\begin{eqnarray}\label{m4}
C^{(S)}_{sum}=\frac{1}{2}\mathrm{log}_{2}\left(1+\underbrace{\mathrm{min}\left\{\gamma_{U_{1}}^{(x_{1})},\gamma_{U_{2}}^{(S,
x_{1})},\gamma_{U_3}^{\left(S,x_{1}\right)}\right\}}_{\mathcal{S}_1}\right)
+\frac{1}{2}\mathrm{log}_{2}\left(1+\underbrace{\mathrm{min}\left\{\gamma_{U_{1}}^{(x_{2})},\gamma_{U_{2}}^{(S,x_{2})},\gamma_{U_3}^{(S,x_{2})}\right\}}_{\mathcal{S}_2}\right)\notag\\
+\frac{1}{2}\mathrm{log}_{2}\left(1+\underbrace{\mathrm{min}\left\{\gamma_{U_{2}}^{(S,x_{r})},\gamma_{U_3}^{(S,x_{r})}\right\}}_{\mathcal{S}_3}\right),
\end{eqnarray}
where $\frac{1}{2}$ results from the two-phase transmission.

\subsection{MRC Decoding Scheme}

In this scheme, realizing that the performance of the achievable
rate is limited by a poor channel, the UE2 and UE3 will not
immediately decode the receptions but instead conserve them until
the second phase comes. To joint jointly decode $x_{1}$ and
$x_{2}$ by utilizing MRC at the UE2 and UE3, the corresponding
effective SNRs for $x_{1}$, $x_{2}$ and $x_{r}$ are respectively
given as
\begin{eqnarray}\label{v2}
\gamma_{U_l}^{\left(M,x_{1}\right)}=\frac{a_1\left|h_{SU_l}\right|^{2}\rho}{a_2\left|h_{SU_l}\right|^{2}\rho+1},
\end{eqnarray},
\begin{eqnarray}\label{v2}
\gamma_{U_l}^{\left(M,x_{2}\right)}=\frac{b_1\left|h_{RU_{l}}\right|^{2}\rho}{b_2\left|h_{RU_{l}}\right|^{2}\rho+1}+a_2\left|h_{SU_{l}}\right|^{2}\rho,
\end{eqnarray}
and
\begin{eqnarray}\label{v3}
\gamma_{U_l}^{\left(M,x_{r}\right)}=b_2\left|h_{RU_{l}}\right|^{2}\rho,
\end{eqnarray}
for $l\in\{2,3\}$. Therefore, in this scheme, the achievable SR
for the proposed capacity enhanced D2D-NOMA system can be obtained
from:
\begin{eqnarray}\label{v4}
C_{sum}^{(M)}=\frac{1}{2}\mathrm{log}_{2}\left(1+\underbrace{\mathrm{min}\left\{\gamma_{U_{1}}^{(x_{1})},\gamma_{U_{2}}^{(M,x_{1})},\gamma_{U_3}^{\left(M,x_{1}\right)}\right\}}_{\mathcal{M}_1}\right)
+\frac{1}{2}\mathrm{log}_{2}\left(1+\underbrace{\mathrm{min}\left\{\gamma_{U_{1}}^{(x_{2})},\gamma_{U_{2}}^{(M,x_{2})},\gamma_{U_3}^{(M,x_{2})}\right\}}_{\mathcal{M}_2}\right)\notag\\
+\frac{1}{2}\mathrm{log}_{2}\left(1+\underbrace{\mathrm{min}\left\{\gamma_{U_{2}}^{(M,x_{r})},\gamma_{U_3}^{(M,x_{r})}\right\}}_{\mathcal{M}_3}\right).
\end{eqnarray}

\section{Ergodic Rate Analysis for the Proposed Two Decoding Strategies}

In this section, we focus on the performance analysis of the
proposed capacity enhanced cooperative D2D-NOMA system for both of
two decoding strategies, which includes the achievable ergodic SR,
outage probability and outage capacity over the Rayleigh fading
channel.

\subsection{Single Signal Decoding Scheme}

Letting $\beta_{SU_i}\triangleq\left|h_{SU_i}\right|^2$, and
$\beta_{RU_l}\triangleq\left|h_{RU_l}\right|^2$, we have
\begin{eqnarray}\label{v6}
\mathcal{S}_1=\min\left\{\frac{a_{1}\beta_{SU_1}\rho}{a_{2}\beta_{SU_1}\rho+1},\frac{a_{1}\beta_{SU_2}\rho}{a_{2}\beta_{SU_2}\rho+1},\frac{a_{1}\beta_{SU_3}\rho}{a_{2}\beta_{SU_3}\rho+1}\right\}.
\end{eqnarray}
Based on \eqref{v6}, the complementary cumulative distribution
function (CCDF) of $\mathcal{S}_1$ can be obtained as
\begin{eqnarray}\label{xx10}
\overline{F}_{\mathcal{S}_1}(s_1)=\Pr\left\{\frac{a_{1}\beta_{SU_1}\rho}{a_{2}\beta_{SU_1}\rho+1}>s_1,\frac{a_{1}\beta_{SU_2}\rho}{a_{2}\beta_{SU_2}\rho+1}>s_1,\frac{a_{1}\beta_{SU_3}\rho}{a_{2}\beta_{SU_3}\rho+1}>s_1\right\}.
\end{eqnarray}
Note that the CCDF of
$\beta_{\delta}=e^{-\frac{x}{\alpha_{\delta}}}$, for
$\delta\in\{SU_{i}, RU_{l}\}$. When $s_1<\frac{a_1}{a_2}$,
\eqref{xx10} can be equivalently represented as
\begin{eqnarray}\label{xx11}
\overline{F}_{\mathcal{S}_1}(s_1)&=&\overline{F}_{SU_{1}}\left(\frac{s_1}{a_{1}\rho-a_{2}\rho
x}\right)\overline{F}_{SU_{2}}\left(\frac{s_1}{a_{1}\rho-a_{2}\rho
x}\right)
\overline{F}_{SU_{3}}\left(\frac{s_1}{a_{1}\rho-a_{2}\rho
x}\right)\notag\\
&=&e^{-\frac{s_1}{a_{1}\rho-a_{2}\rho
s_1}\left(\frac{1}{\alpha_{SU_{1}}}+\frac{1}{\alpha_{SU_{2}}}+\frac{1}{\alpha_{SU_{3}}}\right)}.
\end{eqnarray}
For the case $s_1>\frac{a_1}{a_2}$,
$\overline{F}_{\mathcal{S}_1}(s_1)=0$ always holds due to
\begin{eqnarray}\label{xx12}
\frac{\beta_{SU_{i}}a_{1}\rho}{\beta_{SU_{i}}a_{2}\rho+1}<\frac{a_{1}}{a_{2}}.
\end{eqnarray}
Taking derivative of \eqref{xx11}, the PDF of $\mathcal{S}_1$ is
given as
\begin{eqnarray}\label{xx13}
f_{\mathcal{S}_1}(s_1)=\frac{s_1}{a_{1}\rho-a_{2}\rho
s_1}\left(\frac{1}{\alpha_{SU_{1}}}+\frac{1}{\alpha_{SU_{2}}}+\frac{1}{\alpha_{SU_{3}}}\right)e^{-\frac{s_1}{a_{1}\rho-a_{2}\rho
s_1}\left(\frac{1}{\alpha_{SU_{1}}}+\frac{1}{\alpha_{SU_{2}}}+\frac{1}{\alpha_{SU_{3}}}\right)}.
\end{eqnarray}
Using the equality
\begin{eqnarray}\label{bbb5}
\int_0^{\infty}\mathrm{log}_2\left(1+x\right)f_{X}(x)dx=\frac{1}{2\mathrm{ln}2}\int_0^{\infty}\frac{1-F(x)}{1+x}dx,
\end{eqnarray}
and considering the high SNR case, from \eqref{xx11}, the
achievable ergodic rate for $x_{1}$ can be obtained as follows:
\begin{eqnarray}\label{nn2}
C_{\mathcal{S}_1}^{(x_1)}&=&
\int_{0}^{\frac{a_1}{a_2}}\frac{1}{2}\mathrm{log}_{2}\left(1+s_1\right)dF_{\mathcal{S}_1}(s_1)
+\frac{1}{2}\mathrm{log}_{2}\left(1+\frac{a_1}{a_2}\right)\left(1-F_{\mathcal{S}_1}\left(\frac{a_1}{a_2}\right)\right)\notag\\
&=&\frac{1}{2}\mathrm{log}_{2}\left(1+\frac{a_1}{a_2}\right)-\frac{1}{2\mathrm{ln}2}\int_{0}^{\frac{a_1}{a_2}}\frac{1}{1+s_1}\left(1-e^{-\frac{s_1}{a_{1}\rho-a_{2}\rho
s_1}\left(\frac{1}{\alpha_{SU_{1}}}+\frac{1}{\alpha_{SU_{2}}}+\frac{1}{\alpha_{SU_{3}}}\right)}\right)ds_1\notag\\
&=&\frac{1}{2\mathrm{ln}2}\int_{0}^{\frac{a_1}{a_2}}\frac{e^{-\frac{s_1}{a_{1}\rho-a_{2}\rho
s_1}\left(\frac{1}{\alpha_{SU_{1}}}+\frac{1}{\alpha_{SU_{2}}}+\frac{1}{\alpha_{SU_{3}}}\right)}}{1+s_1}ds_1\notag\\
&=&-\frac{1}{2\mathrm{ln}2}\Bigg(e^{\frac{1}{\rho}\left(\frac{1}{\alpha_{SU_{1}}}+\frac{1}{\alpha_{SU_{2}}}+\frac{1}{\alpha_{SU_{3}}}\right)}\left[\mathrm{Ei}\left(-\frac{1}{\rho}\left(\frac{1}{\alpha_{SU_{1}}}+\frac{1}{\alpha_{SU_{2}}}+\frac{1}{\alpha_{SU_{3}}}\right)\right)\right]\notag\\
&&-e^{\frac{1}{a_2\rho}\left(\frac{1}{\alpha_{SU_{1}}}+\frac{1}{\alpha_{SU_{2}}}+\frac{1}{\alpha_{SU_{3}}}\right)}\left[\mathrm{Ei}\left(-\frac{1}{a_2\rho}\left(\frac{1}{\alpha_{SU_{1}}}+\frac{1}{\alpha_{SU_{2}}}+\frac{1}{\alpha_{SU_{3}}}\right)\right)\right]\Bigg),
\end{eqnarray}
where the third term in \eqref{nn2} is simplified by using
\begin{eqnarray}
\int_{0}^{u}\frac{e^{-\mu
x}dx}{x+\beta}=e^{\mu\beta}\left[\mathrm{Ei}\left(-\mu
u-\mu\beta\right)-\mathrm{Ei}\left(-\mu\beta\right)\right]
\end{eqnarray}
\cite[Eq. (3.352.1)]{3}.

For $\mathcal{S}_{2}$, the corresponding CCDF can be obtained from
\begin{eqnarray}\label{pp1}
\overline{F}_{\mathcal{S}_2}(s_2)&=&\Pr\left\{a_2\beta_{SU_1}\rho>s_2,\frac{b_{1}\beta_{RU_2}\rho}{b_{2}\beta_{RU_2}\rho+1}>s_2,\frac{b_{1}\beta_{RU_3}\rho}{b_{2}\beta_{RU_3}\rho+1}>s_2\right\}\notag\\
&=&\overline{F}_{SU_{1}}\left(\frac{s_2}{a_{2}\rho
}\right)\overline{F}_{RU_{2}}\left(\frac{s_2}{b_{1}\rho-b_{2}\rho
s_2}\right)
\overline{F}_{RU_{3}}\left(\frac{s_2}{b_{1}\rho-b_{2}\rho
s_2}\right\}\notag\\
&=&e^{-\frac{s_2}{b_{1}\rho-b_{2}\rho
s_2}\left(\frac{1}{\alpha_{RU_{2}}}+\frac{1}{\alpha_{RU_{3}}}\right)-\frac{s_2}{a_2\rho\alpha_{SU_{1}}}}.
\end{eqnarray}
Similar to \cite{h1}, considering a high transmit SNR case, which
leads to
\begin{eqnarray}
\frac{b_1\left|h_{RU_l}\right|^{2}\rho}{b_2\left|h_{RU_l}\right|^{2}\rho+1}\sim\frac{b1}{b2},
\end{eqnarray}
the effective capacity of $x_2$ can be approximately obtained from
\begin{eqnarray}\label{nn3}
C_{\mathcal{S}_2}^{(x_2)}&\sim&\frac{e^{\frac{1}{a_2\rho\alpha_{SU_1}}}}{2\mathrm{ln}2}\left(\mathrm{Ei}\left(-\frac{1}{a_2\rho\alpha_{SU_1}}\left(1+\frac{b1}{b2}\right)\right)-\mathrm{Ei}\left(-\frac{1}{a_2\rho\alpha_{SU_1}}\right)\right)
\end{eqnarray}
Correspondingly, for $\mathcal{S}_{3}$, the corresponding CCDF is
given as
\begin{eqnarray}\label{pp2}
\overline{F}_{\mathcal{S}_3}(s_3)&=&\Pr\left\{b_2\beta_{RU_2}\rho>s_3, b_2\beta_{RU_3}\rho>s_3\right\}\notag\\
&=&\overline{F}_{RU_{2}}\left(\frac{s_3}{b_{2}\rho
}\right)\overline{F}_{RU_{3}}\left(\frac{s_3}{b_{2}\rho
}\right)\notag\\
&=&e^{-\frac{s_3}{b_2\rho}\left(\frac{1}{\alpha_{RU_2}}+\frac{1}{\alpha_{RU_3}}\right)},
\end{eqnarray}
with the effective rate as
\begin{eqnarray}\label{pp3}
C_{\mathcal{S}_3}^{(x_3)}=\frac{e^{\frac{1}{b_2\rho}\left(\frac{1}{\alpha_{RU_2}}+\frac{1}{\alpha_{RU_3}}\right)}}{2\mathrm{ln}2}
\mathrm{Ei}\left(-\frac{1}{b_2\rho}\left(\frac{1}{\alpha_{RU_2}}+\frac{1}{\alpha_{RU_3}}\right)\right)
\end{eqnarray}

Synthesizing \eqref{nn2}, \eqref{nn3} and \eqref{pp3}, the
achievable SR for the \emph{single signal decoding scheme} can be
expressed as
\begin{eqnarray}\label{pp4}
C_{\mathcal{S}}^{(sum)}&=&-\frac{1}{2\mathrm{ln}2}\Bigg(e^{\frac{1}{\rho}\left(\frac{1}{\alpha_{SU_{1}}}+\frac{1}{\alpha_{SU_{2}}}+\frac{1}{\alpha_{SU_{3}}}\right)}\left[\mathrm{Ei}\left(-\frac{1}{\rho}\left(\frac{1}{\alpha_{SU_{1}}}+\frac{1}{\alpha_{SU_{2}}}+\frac{1}{\alpha_{SU_{3}}}\right)\right)\right]\notag\\
&&-e^{\frac{1}{a_2\rho}\left(\frac{1}{\alpha_{SU_{1}}}+\frac{1}{\alpha_{SU_{2}}}+\frac{1}{\alpha_{SU_{3}}}\right)}\left[\mathrm{Ei}\left(-\frac{1}{a_2\rho}\left(\frac{1}{\alpha_{SU_{1}}}+\frac{1}{\alpha_{SU_{2}}}+\frac{1}{\alpha_{SU_{3}}}\right)\right)\right]\Bigg)\notag\\
&&+\frac{e^{\frac{1}{a_2\rho\alpha_{SU_1}}}}{2\mathrm{ln}2}\left(\mathrm{Ei}\left(-\frac{1}{a_2b_2\rho\alpha_{SU_1}}\right)-\mathrm{Ei}\left(-\frac{1}{a_2\rho\alpha_{SU_1}}\right)\right)\notag\\
&&+\frac{e^{\frac{1}{b_2\rho}\left(\frac{1}{\alpha_{RU_2}}+\frac{1}{\alpha_{RU_3}}\right)}}{2\mathrm{ln}2}
\mathrm{Ei}\left(-\frac{1}{b_2\rho}\left(\frac{1}{\alpha_{RU_2}}+\frac{1}{\alpha_{RU_3}}\right)\right).
\end{eqnarray}
By further applying the approximations $\mathrm{Ei}(-x)\sim
\mathrm{Ec}+\mathrm{ln}~x$ and $e^x\sim 1+x$, for small $x$, after
some algebra, \eqref{pp4} can be finally rewritten as
\begin{eqnarray}\label{pp5}
C_{\mathcal{S}}^{(sum)}\sim\frac{1}{2\mathrm{ln}2}\bigg(
\frac{\mathcal{A}}{a_2}\left(-a_1\mathrm{Ec}-a_1\mathrm{ln}\mathcal{A}+\mathrm{ln}a_2\right)+\mathrm{ln}a_2+\left(1+\frac{1}{a_2\rho\alpha_{SU_1}}\right)\mathrm{ln}b_2\notag\\
~~~~+\left(1+\mathcal{B}\right)\left(\mathrm{Ec}+\mathrm{ln}\mathcal{B}\right)\bigg),
\end{eqnarray}
where
$\mathcal{A}=\frac{1}{\rho}\left(\frac{1}{\alpha_{SU_{1}}}+\frac{1}{\alpha_{SU_{2}}}+\frac{1}{\alpha_{SU_{3}}}\right)$
and
$\mathcal{B}=\frac{1}{b_2\rho}\left(\frac{1}{\alpha_{RU_2}}+\frac{1}{\alpha_{RU_3}}\right)$.

\subsection{MRC Decoding Scheme}

Since that the final expressions of the ergodic rate for
$\mathcal{M}_1$ and $\mathcal{M}_3$ are equivalent to the ones of
$\mathcal{S}_1$ and $\mathcal{S}_3$, respectively. In this
subsection, we will focus on the closed-form expression of
$\mathcal{M}_2$:
\begin{eqnarray}\label{s1}
\mathcal{M}_2=\min\left\{a_{2}\beta_{SU_1}\rho,\frac{b_{1}\beta_{RU_2}\rho}{b_{2}\beta_{RU_2}\rho+1}+a_2\beta_{SU_2}\rho,\frac{b_{1}\beta_{RU_3}\rho}{b_{2}\beta_{RU_3}\rho+1}+a_2\beta_{SU_3}\rho\right\}
\end{eqnarray}
with the corresponding CCDF as
\begin{eqnarray}\label{pp2}
\overline{F}_{\mathcal{M}_2}(m_2)&=&\Pr\left\{a_{2}\beta_{SU_1}\rho>m_2,\frac{b_{1}\beta_{RU_2}\rho}{b_{2}\beta_{RU_2}\rho+1}+a_2\beta_{SU_2}\rho>m_2,\frac{b_{1}\beta_{RU_3}\rho}{b_{2}\beta_{RU_3}\rho+1}+a_2\beta_{SU_3}\rho>m_2\right\}\notag\\
&=&\overline{F}_{SU_{1}}\left(\frac{m_2}{a_{2}\rho
}\right)\underbrace{\overline{F}_{SU_{2}}\left(\frac{m_2-\frac{b_{1}\beta_{RU_2}\rho}{b_{2}\beta_{RU_2}\rho+1}}{a_2\rho}\right)}_{\mathcal{P}_1}\underbrace{\overline{F}_{SU_{3}}\left(\frac{m_2-\frac{b_{1}\beta_{RU_3}\rho}{b_{2}\beta_{RU_3}\rho+1}}{a_2\rho}\right)}_{\mathcal{P}_2}.
\end{eqnarray}
For $\mathcal{P}_1$, we have the exact expression as following:
\begin{equation}
\mathcal{P}_1\!\!=\!\!\left\{
\begin{array}{rl}
&\!\!\!\!\!\!\Pr\left\{\frac{b_{1}\beta_{RU_2}\rho}{b_{2}\beta_{RU_2}\rho+1}<m_2\right\}\Pr\left\{
a_2\beta_{SU_2}\rho>m_2-\frac{b_{1}\beta_{RU_2}\rho}{b_{2}\beta_{RU_2}\rho+1}\right\}+\Pr\left\{\frac{b_{1}\beta_{RU_2}\rho}{b_{2}\beta_{RU_2}\rho+1}>m_2\right\},~
m_2<\frac{b1}{b2}\\
&\!\!\!\!\!\!\Pr\left\{
a_2\beta_{SU_2}\rho>m_2-\frac{b_{1}\beta_{RU_2}\rho}{b_{2}\beta_{RU_2}\rho+1}\right\},~~~
m_2>\frac{b1}{b2}¡£
\end{array}
\right.
\end{equation}
For $m_2>\frac{b_1}{b_2}$, we have
\begin{equation}\label{y1}
\mathcal{P}_1=\frac{1}{\beta_{RU_2}}e^{-\frac{m_2}{\beta_{SU_2}\rho}}\int_{0}^{\infty}e^{-\frac{1}{\beta_{SU_2}}
\left(\frac{b_1u}{b_2u\rho+1}\right)-\frac{u}{\beta_{RU_2}}}du.
\end{equation}
In the meanwhile that, for $m_2<\frac{b_1}{b_2}$, the effective
CCDF for $\mathcal{M}_2$ can be obtained from
\begin{equation}\label{y2}
\mathcal{P}_1=e^{-\frac{1}{\beta_{RU_2}}\left(\frac{m_2}{b_1\rho-m_2b_2\rho}\right)}+
\frac{1}{\beta_{RU_2}}\left(1-e^{-\frac{1}{\beta_{RU_2}}\left(\frac{m_2}{b_1\rho-m_2b_2\rho}\right)}\right)e^{-\frac{m_2}{\beta_{SU_2}}}
\int_{0}^{\infty}e^{-\frac{1}{\beta_{SU_2}}
\left(\frac{b_1u}{b_2u\rho+1}\right)-\frac{u}{\beta_{RU_2}}}du.
\end{equation}
Clearly, \eqref{y2} is quite involved to derive the exact ergodic
achievable rate. Therefore, we turn to examine it in an
approximate way, i.e.,
$\frac{b_{1}\beta_{RU_2}\rho}{b_{2}\beta_{RU_2}\rho+1}\sim\frac{b_1}{b_2}$,
for a high transmit SNR. By this way, the CCDF for $\mathcal{M}_2$
can be finally obtained as
\begin{equation}\label{y3}
\overline{F}_{\mathcal{M}_2}(m_2)=e^{-\frac{m_2}{a_2\rho}\left(\frac{1}{\beta_{SU_1}}+\frac{1}{\beta_{SU_2}}+\frac{1}{\beta_{SU_3}}\right)+\frac{b_1}{b_2\rho}\left(\frac{1}{\beta_{RU_2}}+\frac{1}{\beta_{RU_3}}\right)},
\end{equation}
with the corresponding ergodic achievable rate as
\begin{eqnarray}\label{y4}
C_{\mathcal{M}_2}^{(x_2)}&=&\frac{1}{2\mathrm{ln}2}\int_{0}^{\infty}\frac{e^{-\frac{m_2}{a_2\rho}\left(\frac{1}{\beta_{SU_1}}+\frac{1}{\beta_{SU_2}}+\frac{1}{\beta_{SU_3}}\right)+\frac{b_1}{b_2\rho}\left(\frac{1}{\beta_{RU_2}}+\frac{1}{\beta_{RU_3}}\right)}}{1+m_2}dm_2\notag\\
&=&\frac{-e^{\frac{1}{a_2\rho}\left(\frac{1}{\beta_{SU_1}}+\frac{1}{\beta_{SU_2}}+\frac{1}{\beta_{SU_3}}\right)}}{2\mathrm{ln}2}
\mathrm{Ei}\left(-\frac{1}{a_2\rho}\left(\frac{1}{\beta_{SU_1}}+\frac{1}{\beta_{SU_2}}+\frac{1}{\beta_{SU_3}}\right)\right),
\end{eqnarray}
where the equation \eqref{bbb5} is used. By substituting
\eqref{nn2}, \eqref{pp3}, and \eqref{y4} back into \eqref{v4}, the
achievable SR for the proposed capacity enhanced D2D-NOMA system
can be expressed as
\begin{eqnarray}\label{y5}
C_{sum}^{(M)}&=&\frac{-e^{\frac{1}{a_2\rho}\left(\frac{1}{\beta_{SU_1}}+\frac{1}{\beta_{SU_2}}+\frac{1}{\beta_{SU_3}}\right)}}{2\mathrm{ln}2}
\mathrm{Ei}\left(-\frac{1}{a_2\rho}\left(\frac{1}{\beta_{SU_1}}+\frac{1}{\beta_{SU_2}}+\frac{1}{\beta_{SU_3}}\right)\right)\notag\\
&&-\frac{1}{2\mathrm{ln}2}\Bigg(e^{\frac{1}{\rho}\left(\frac{1}{\alpha_{SU_{1}}}+\frac{1}{\alpha_{SU_{2}}}+\frac{1}{\alpha_{SU_{3}}}\right)}\left[\mathrm{Ei}\left(-\frac{1}{\rho}\left(\frac{1}{\alpha_{SU_{1}}}+\frac{1}{\alpha_{SU_{2}}}+\frac{1}{\alpha_{SU_{3}}}\right)\right)\right]\notag\\
&&-\frac{e^{\frac{1}{a_2\rho}\left(\frac{1}{\alpha_{SU_{1}}}+\frac{1}{\alpha_{SU_{2}}}+\frac{1}{\alpha_{SU_{3}}}\right)}}{2\mathrm{ln}2}\left[\mathrm{Ei}\left(-\frac{1}{a_2\rho}\left(\frac{1}{\alpha_{SU_{1}}}+\frac{1}{\alpha_{SU_{2}}}+\frac{1}{\alpha_{SU_{3}}}\right)\right)\right]\Bigg)\notag\\
&&+\frac{e^{\frac{1}{b_2\rho}\left(\frac{1}{\alpha_{RU_2}}+\frac{1}{\alpha_{RU_3}}\right)}}{2\mathrm{ln}2}
\mathrm{Ei}\left(-\frac{1}{b_2\rho}\left(\frac{1}{\alpha_{RU_2}}+\frac{1}{\alpha_{RU_3}}\right)\right).
\end{eqnarray}
Similarly, employing the approximations $\mathrm{Ei}(-x)\sim
\mathrm{Ec}+\mathrm{ln}~x$ and $e^x\sim 1+x$, for small $x$,
\eqref{y5} can be approximately rewritten into
\begin{eqnarray}\label{pp5}
C_{\mathcal{S}}^{(sum)}\sim\frac{1}{2\mathrm{ln}2}\bigg(
\frac{\mathcal{A}}{a_2}\left(-\left(a_1+1\right)\mathrm{Ec}-a_1\mathrm{ln}\mathcal{A}+\mathrm{ln}\frac{a_2^2}{\mathcal{A}}\right)+\mathrm{ln}a_2
+\mathcal{B}\left(\mathrm{Ec}+\mathrm{ln}\mathcal{B}\right)+\mathrm{ln}\frac{a_2\mathcal{B}}{\mathcal{A}}\bigg).
\end{eqnarray}
\emph{Discussion:} It is clear, for these two proposed decoding
strategies, the difference of the sum-rate is provided by
$\mathcal{S}_2$ and $\mathcal{M}_2$, which are the corresponding
received SNR for $x_2$. More specifically, for a high transmit SNR
case, the first term in \eqref{pp1} and \eqref{s1} will be greater
than the last two terms, since that the distance between the BS
and relay is normal smaller than the one between the BS and users.
With this observation, comparing the last two terms in \eqref{pp1}
and \eqref{s1}, it is easy to see that, the difference
``$a_2\beta_{SU_2}\rho$" and ``$a_2\beta_{SU_3}\rho$" will result
in a significant performance improvement for the proposed
\emph{MRC decoding scheme}. In the following section, the
numerical results will be used to further confirm the correctness
of our analysis.

\section{Outage Probability and Outage Capacity Analysis}

\subsection{Outage Probability Analysis}

In this subsection, we will derive the outage probability into
asymptotic expressions for the proposed D2D-NOMA system with two
decoding schemes. Assuming $\mathcal{R}^{x_{1}}_{\mathcal{T}}$,
$\mathcal{R}^{x_{2}}_{\mathcal{T}}$ and
$\mathcal{R}^{x_{r}}_{\mathcal{T}}$ denoted as the predefined
target rate thresholds of the symbols $x_{1}$, $x_{2}$ and
$x_{3}$, respectively, the outage event occurs when
$\mathcal{R}_{x_{1}}$, $\mathcal{R}_{x_{2}}$ and
$\mathcal{R}_{x_{r}}$ are smaller than that of the
$\mathcal{R}^{x_{1}}_{\mathcal{T}}$,
$\mathcal{R}^{x_{2}}_{\mathcal{T}}$ and
$\mathcal{R}^{x_{r}}_{\mathcal{T}}$, where
$\mathcal{R}^{x_{g}}_{\mathcal{T}}$ denotes the achievable rate
for the date $x_{g}$, for $g\in\left\{1,2,r\right\}$.

The exact outage probability for the \emph{MRC decoding scheme}
can be written as:
\begin{eqnarray}\label{16}
P_{\mathcal{S}}&=&1-\mathrm{Pr}\left[\mathcal{R}_{x_{1}}>\mathcal{R}^{x_{1}}_{\mathcal{T}},
\mathcal{R}_{x_{2}}>\mathcal{R}^{x_{2}}_{\mathcal{T}},
\mathcal{R}_{x_{r}}>\mathcal{R}^{x_{r}}_{\mathcal{T}}\right]\notag\\
&=&1-\mathrm{Pr}\left\{\underbrace{\left(\mathcal{M}_1>2^{2\mathcal{R}^{x_{1}}_{\mathcal{T}}}-1\right)}_{\mathcal{K}_1}\cap
\underbrace{\left(\mathcal{M}_2>2^{2\mathcal{R}^{x_{2}}_{\mathcal{T}}}-1\right)}_{\mathcal{K}_2}\cap
\underbrace{\left(\mathcal{M}_3>2^{2\mathcal{R}^{x_{r}}_{\mathcal{T}}}-1\right)}_{\mathcal{K}_r}\right\}.
\end{eqnarray}
Further assuming
$\mathcal{W}_1=2^{\mathcal{R}^{x_{1}}_{\mathcal{T}}}-1$,
$\mathcal{W}_2=2^{\mathcal{R}^{x_{2}}_{\mathcal{T}}}-1$ and
$\mathcal{W}_r=2^{\mathcal{R}^{x_{r}}_{\mathcal{T}}}-1$,
respectively, the final closed-form expressions of
$\mathcal{K}_1$, $\mathcal{K}_2$ and $\mathcal{K}_r$ can be
obtained as:
\begin{eqnarray}\label{17}
\mathcal{K}_1&=&\Pr\left\{\min\left(\frac{a_{1}\beta_{SU_1}\rho}{a_{2}\beta_{SU_1}\rho+1},\frac{a_{1}\beta_{SU_2}\rho}{a_{2}\beta_{SU_2}\rho+1},\frac{a_{1}\beta_{SU_3}\rho}{a_{2}\beta_{SU_3}\rho+1}\right)
>\mathcal{W}_1\right\}\notag\\
&=&\Pr\left\{\frac{a_{1}\beta_{SU_1}\rho}{a_{2}\beta_{SU_1}\rho+1}>\mathcal{W}_t\right\}\Pr\left\{\frac{a_{1}\beta_{SU_2}\rho}{a_{2}\beta_{SU_2}\rho+1}>\mathcal{W}_1\right\}\Pr\left\{\frac{a_{1}\beta_{SU_3}\rho}{a_{2}\beta_{SU_3}\rho+1}>\mathcal{W}_1\right\}\notag\\
&=&\left\{
\begin{array}{l}
e^{-\frac{\mathcal{W}_1}{a_{1}\rho-a_{2}\rho
\mathcal{W}_1}\left(\frac{1}{\alpha_{SU_{1}}}+\frac{1}{\alpha_{SU_{2}}}+\frac{1}{\alpha_{SU_{3}}}\right)}, ~~\mathrm{for}~~\mathcal{W}_1<\frac{a_{1}}{a_{2}},\\
0~~~~~~~~~~~~,
~~~~~~~~~~~~~~~~~~~~~~~\mathrm{for}~~\mathcal{W}_1>\frac{a_{1}}{a_{2}},
\end{array}
\right.
\end{eqnarray}
\begin{eqnarray}\label{18}
\mathcal{K}_2&=&\Pr\left\{\mathcal{M}_2>2^{2\mathcal{R}^{x_{2}}_{\mathcal{T}}}-1\right\}\notag\\
&=&\Pr\left\{\min\left\{a_{2}\beta_{SU_1}\rho,\frac{b_{1}\beta_{RU_2}\rho}{b_{2}\beta_{RU_2}\rho+1}+a_2\beta_{SU_2}\rho,\frac{b_{1}\beta_{RU_3}\rho}{b_{2}\beta_{RU_3}\rho+1}+a_2\beta_{SU_3}\rho\right\}
>\mathcal{W}_2\right\}\notag\\
&=&\Pr\left\{a_{2}\beta_{SU_1}\rho>\mathcal{M}_2\right\}\Pr\left\{\frac{b_{1}\beta_{RU_2}\rho}{b_{2}\beta_{RU_2}\rho+1}+a_2\beta_{SU_2}\rho>\mathcal{M}_2\right\}\Pr\left\{\frac{b_{1}\beta_{RU_3}\rho}{b_{2}\beta_{RU_3}\rho+1}+a_2\beta_{SU_3}\rho>\mathcal{M}_2\right\}\notag\\
&=&\left\{
\begin{array}{l}
e^{-\frac{\mathcal{W}_2}{a_2\rho}\left(\frac{1}{\beta_{SU_1}}+\frac{1}{\beta_{SU_2}}+\frac{1}{\beta_{SU_3}}\right)+\frac{b_1}{b_2\rho}\left(\frac{1}{\beta_{RU_2}}+\frac{1}{\beta_{RU_3}}\right)},~~\mathrm{for}~~\mathcal{W}_2<\frac{b_1}{b_2},\\
0~~~~~~~~~~, ~~\mathrm{for}~~\mathcal{W}_2>\frac{b_1}{b_2},
\end{array}
\right.
\end{eqnarray}
and
\begin{eqnarray}\label{19}
\mathcal{K}_r&=&\Pr\left\{\left(b_2\beta_{RU_2}\rho, b_2\beta_{RU_3}\rho\right)>\mathcal{W}_r\right\}\notag\\
&=&\Pr\left\{b_2\beta_{RU_2}\rho>\mathcal{W}_r\right\}\Pr\left\{b_2\beta_{RU_3}\rho>\mathcal{W}_r\right\}\notag\\
&=&e^{-\frac{\mathcal{W}_r}{b_2\rho}\left(\frac{1}{\alpha_{RU_2}}+\frac{1}{\alpha_{RU_3}}\right)}.
\end{eqnarray}
With the conditions $
\left\{\mathcal{W}_1,\mathcal{W}_2\right\}<\min\left\{\frac{a_{1}}{a_{2}},\frac{b_{1}}{b_{2}}\right\},$
substituting \eqref{17}, \eqref{18} and \eqref{19} back into
\eqref{16}, we have the outage probability in a closed-form
expression as
\begin{eqnarray}\label{tt2}
P_{\mathcal{M}}\sim1-e^{-\frac{\mathcal{W}_1}{a_{1}\rho-a_{2}\rho
\mathcal{W}_1}\left(\frac{1}{\alpha_{SU_{1}}}+\frac{1}{\alpha_{SU_{2}}}+\frac{1}{\alpha_{SU_{3}}}\right)
-\frac{\mathcal{W}_r}{b_2\rho}\left(\frac{1}{\alpha_{RU_2}}+\frac{1}{\alpha_{RU_3}}\right)
-\frac{\mathcal{W}_2-\frac{b_1}{b_2}}{a_2\rho}\left(\frac{1}{\beta_{SU_1}}+\frac{1}{\beta_{SU_2}}+\frac{1}{\beta_{SU_3}}\right)}.\notag\\
\end{eqnarray}

\subsection{Outage Capacity Analysis}

In this subsection, the outage capacity analysis of our proposed
D2D-NOMA system is analyzed. Since it is quit involved to obtain
the exact outage capacity for our proposed scheme, in this paper,
we turn to find an approximate way to derive the outage capacity,
i.e., consider the high transmit SNR case.

Employing the approximation $e^x\sim1+x$ for small $x$ and letting
$P_{\mathcal{M}}=\varepsilon_{\mathcal{M}}$, \eqref{tt2} can be
approximately rewritten into:
\begin{eqnarray}\label{21t}
\varepsilon_{\mathcal{M}}\sim
\frac{\mathcal{W}_r}{b_2\rho}\left(\frac{1}{\alpha_{RU_2}}+\frac{1}{\alpha_{RU_3}}\right)+\frac{\mathcal{W}_2-\frac{b_1}{b_2}}{a_2\rho}\left(\frac{1}{\beta_{SU_1}}+\frac{1}{\beta_{SU_2}}+\frac{1}{\beta_{SU_3}}\right)
.
\end{eqnarray}
It is worth mentioning that, in a high transmit SNR case, for the
condition $\mathcal{W}_1<\frac{a_1}{a_2}$,
$\mathcal{K}_1\triangleq1$ always holds. Therefore, to satisfy the
QoS condition, setting the target thresholds as
$\mathcal{W}_1=\mathcal{W}_2=\mathcal{W}_r=\mathcal{W}_{\mathcal{S}}=\mathcal{W}_{\mathcal{M}}$,
\eqref{21t} can be finally reexpressed as
\begin{eqnarray}\label{23t}
\mathcal{W}_{\mathcal{M}}=
\Bigg(\frac{1}{b_2\rho}\left(\frac{1}{\alpha_{RU_2}}+\frac{1}{\alpha_{RU_3}}\right)
+\frac{1}{a_2\rho}\left(\frac{1}{\beta_{SU_1}}+\frac{1}{\beta_{SU_2}}
+\frac{1}{\beta_{SU_3}}\right)\Bigg)^{-1}\notag\\
\times\left(\varepsilon_{\mathcal{M}}
+\frac{b_1}{a_2b_2\rho}\left(\frac{1}{\beta_{SU_1}}+\frac{1}{\beta_{SU_2}}+\frac{1}{\beta_{SU_3}}\right)\right),
\end{eqnarray}
with the outage capacity:
\begin{eqnarray}\label{24t}
C_{out}^{\mathcal{M}}=\frac{1}{2}\mathrm{log}_2\left(1+\mathcal{W}_{\mathcal{M}}\right).
\end{eqnarray}

\section{Numerical Result}

In this section, the performance of our proposed D2D-NOMA schemes
in terms of the ergodic SR, outage probability and outage capacity
are evaluated by using computer simulations and comparing to the
benchmarks. All the numerical results are averaged over $100,000$
channel realizations. In the following results, ``Simulation" and
``Analysis" are used to denote the simulation and analytical
results, respectively. In the meanwhile, the ``Recent D2D work",
``Enhance-D2D Single", and ``Enhance-D2D MRC" denote the
convention NOMA scheme in \cite{a14}, \emph{single signal decoding
scheme} and \emph{MRC decoding scheme}, respectively.

\begin{figure}
\begin{center}
\includegraphics [width=110mm]{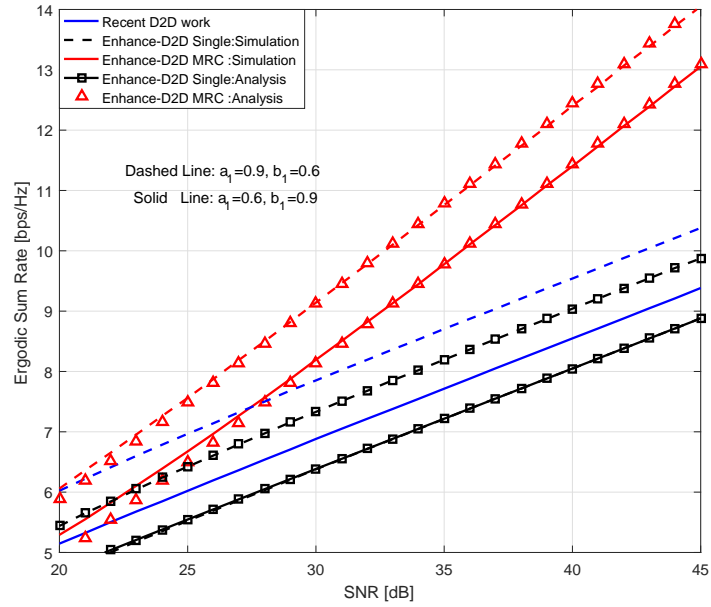}\\
\caption{\label{9} The ergodic SRs achieved by our proposed
\emph{Single Decoding scheme}, \emph{MRC Decoding scheme}, and the
\emph{Recent D2D scheme} with fixed $\alpha_{SU_1}=5$,
$\alpha_{SU_2}=1$, $\alpha_{SU_3}=1$, $\alpha_{RU_2}=2$,
$\alpha_{RU_3}=10$, and $a_1=\{0.6, 0.9\}$, $b_1=\{0.9, 0.6\}$
versus the transmission SNR.}
\end{center}
\end{figure}

\begin{figure}
\begin{center}
\includegraphics [width=110mm]{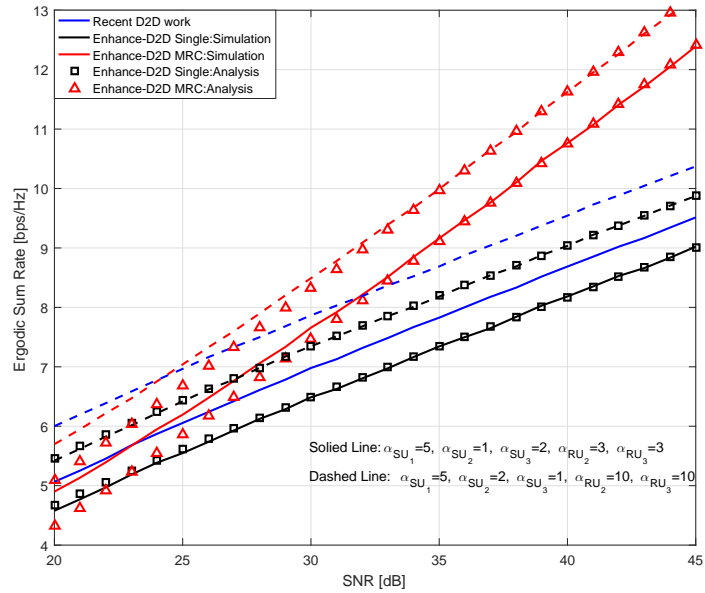}\\
\caption{\label{9} The ergodic SRs achieved by our proposed
\emph{Single Decoding scheme}, \emph{MRC Decoding scheme}, and the
\emph{Recent D2D scheme} with fixed $\alpha_{SU_1}=5$,
$\alpha_{SU_2}=\left\{1,2\right\}$,
$\alpha_{SU_3}=\left\{2,1\right\}$,
$\alpha_{RU_2}=\left\{3,10\right\}$,
$\alpha_{RU_3}=\left\{3,10\right\}$, and $a_1=0.9$, $b_1=0.6$
versus the transmit SNR.}
\end{center}
\end{figure}

Figs. 2 and 3 illustrate how the transmit SNR at the BS affects
the ergodic SR performance of our proposed \emph{single signal
decoding scheme}, \emph{MRC decoding scheme} and the conventional
D2D-NOMA scheme in \cite{a14}. In Fig. 2, we fixed
$\alpha_{SU_1}=5$, $\alpha_{SU_2}=1$, $\alpha_{SU_3}=1$,
$\alpha_{RU_2}=2$, $\alpha_{RU_3}=10$, for two power allocation
setting groups as $\{a_1=0.9, b_1=0.6\}$ and $\{a_1=0.6,
b_1=0.9\}$. In the meanwhile, we set $\alpha_{SU_1}=5$,
$\alpha_{SU_2}=\{1, 2\}$, $\alpha_{SU_3}=\{2, 1\}$,
$\alpha_{RU_2}=\{3, 10\}$, $\alpha_{RU_3}=\{10, 3\}$ with
$\{a_1=0.9, b_1=0.6\}$ for Fig. 3, respectively. As can be
observed from both Figs. 2 and 3, with a increasing transmit SNR,
the ergodic SR increases for all the schemes. Specifically, the
proposed \emph{MRC decoding scheme} overwhelms the \emph{single
signal decoding scheme} and conventional D2D-NOMA one, and it is
worth pointing out that the analytical results match the
simulation results well for both of the proposed two schemes
especially at the high SNR region, which validate our theoretical
analysis in Section III. Remarkably, the ergodic SR of the
proposed \emph{single signal decoding scheme} is worse than that
of the convention D2D-NOMA one. It is reasonable, because the
additional channel $h_{SU_3}$ between the BS and UE3 is normal a
week channel, which limits the corresponding received SNR for
$x_2$. Clearly, the proposed \emph{MRC decoding scheme}
effectively refrain from this negative influence. Moreover, it is
also shown that, with a greater power allocation factor $a_1$, and
channels coefficients $\alpha_{SU_2}$, $\alpha_{RU_2}$,
$\alpha_{RU_3}$, the ergodic SR for all the schemes are also
improved.

\begin{figure}
\begin{center}
\includegraphics [width=110mm]{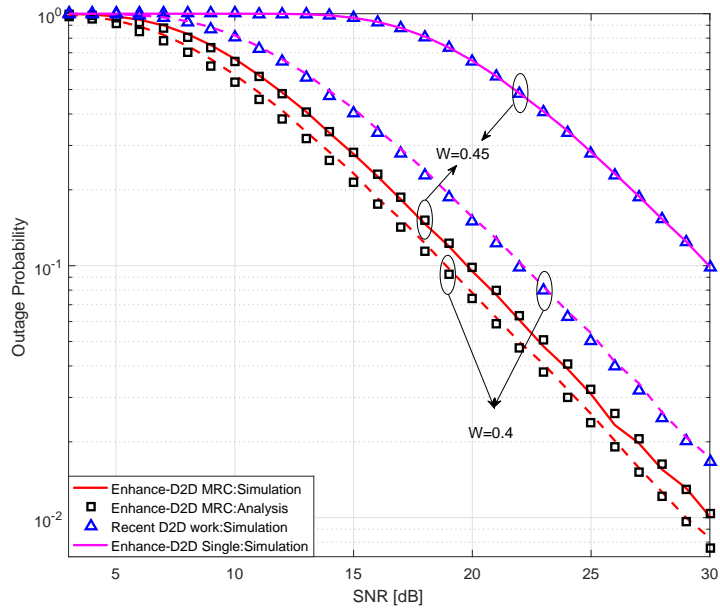}\\
\caption{\label{9} The outage probability for our proposed
\emph{Single Decoding scheme}, \emph{MRC Decoding scheme}, and the
\emph{Recent D2D scheme} with fixed $\alpha_{SU_1}=5$,
$\alpha_{SU_1}=1$, $\alpha_{SU_3}=2$, $\alpha_{RU_2}=1$,
$\alpha_{RU_3}=10$, and $a_1=0.9$, $b_1=0.6$ with the target rate
as 0.4 and 0.45 versus the transmit SNR.}
\end{center}
\end{figure}

Fig. 4 demonstrates the performance of the outage probability for
the proposed schemes and conventional one in both simulation and
analytical results with QoS thresholds as
$\mathcal{W}_1=\mathcal{W}_2=\mathcal{W}_3=\left\{0.4,
0.45\right\}$, where the power allocation factors are set as
$a_1=0.9$, and $b_1=0.6$, while that $\alpha_{SU_1}=5$,
$\alpha_{SU_1}=1$, $\alpha_{SU_3}=2$, $\alpha_{RU_2}=1$,
$\alpha_{RU_3}=10$. In Fig. 4, it can be seen that the
probabilities decrease with a increasing transmit SNR.
Particularly, the increase of QoS thresholds improves the above
outage probabilities dramatically. The asymptotic simulations are
also provided to confirm the close agreement between he simulation
and analytical results. Moreover, the proposed \emph{MRC decoding
scheme} outperforms the other two schemes significantly, since
that our proposed \emph{MRC decoding scheme} provides a greater
corresponding received SNR. In addition, it is also observed that
the outage probability are similar for the \emph{single signal
decoding scheme} and convention D2D-NOMA one.

\begin{figure}
\begin{center}
\includegraphics [width=110mm]{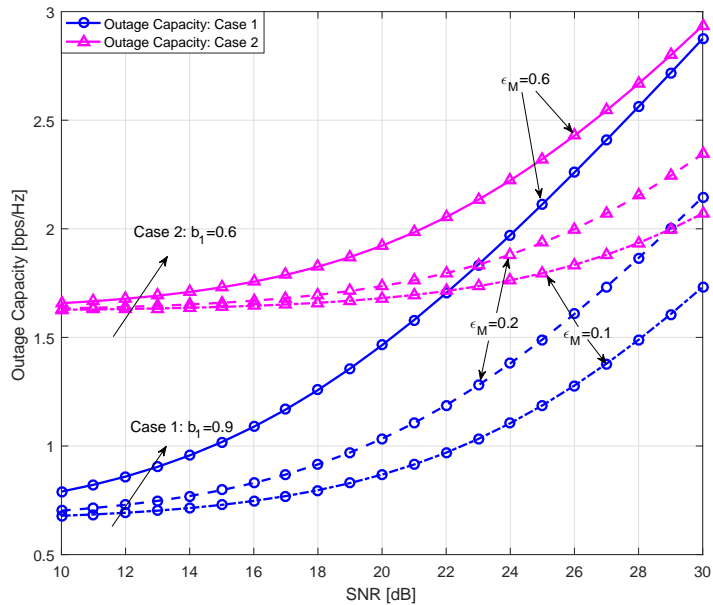}\\
\caption{\label{9} The outage capacity of our proposed \emph{MRC
Decoding scheme} with fixed $\alpha_{SU_1}=20$, $\alpha_{SU_1}=1$,
$\alpha_{SU_3}=10$, $\alpha_{RU_2}=25$, $\alpha_{RU_3}=30$, and
$a_1=0.9$, $b_1=\{0.6, 0.9\}$, $\varepsilon_{\mathcal{M}}=\{0.1,
0.2, 0.6\}$ with respect to the transmit SNR.}
\end{center}
\end{figure}

In Fig. 5, we present the outage capacity with respect to the
transmit SNR and $\varepsilon$, where $\varepsilon$ is set to be
$\left\{0.1, 0.2, 0.6\right\}$ with fixed $\alpha_{SU_1}=20$,
$\alpha_{SU_1}=1$, $\alpha_{SU_3}=10$, $\alpha_{RU_2}=25$,
$\alpha_{RU_3}=30$, and the power allocation factors as
$a_1=0.9$, $b_1=\{0.6, 0.9\}$, respectively. It is clear, the
outage sum capacity decreases with a increasing value of
$\varepsilon$. More specifically, it is noted that for any given
power allocation factor $b_1$, the outage sum capacity is better
for a lower $b_1$ case.

\begin{figure}
\begin{center}
\includegraphics [width=110mm]{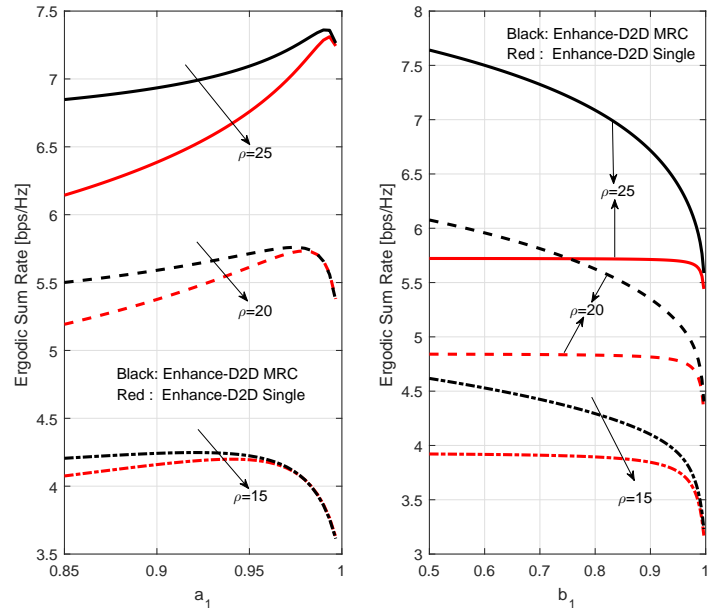}\\
\caption{\label{9} The ergodic SRs achieved by our proposed
\emph{Single Decoding scheme} and \emph{MRC Decoding scheme} with
fixed $\alpha_{SU_1}=5$, $\alpha_{SU_2}=1$, $\alpha_{SU_3}=1$,
$\alpha_{RU_2}=10$, $\alpha_{RU_3}=10$ and $\rho=\left\{15, 20,
25\right\} \mathrm{dB}$ versus the power allocation factors $a_1$
and $b_1$.}
\end{center}
\end{figure}

Fig. 6 plots the impact of the power allocation factors $a_1$ as
well as $b_1$ on the performance of the ergodic SR for the
\emph{single signal decoding scheme} and \emph{MRC decoding
scheme}, respectively. For both figures, we fixed
$\alpha_{SU_1}=5$, $\alpha_{SU_2}=1$, $\alpha_{SU_3}=1$,
$\alpha_{RU_2}=10$, $\alpha_{RU_3}=10$ for different transmit SNRs
as $\rho=\left\{15, 20, 25\right\}~\mathrm{dB}$. Each subfigure in
Fig. 6 considers fixing one power allocation factor, i.e., in Fig.
6(a), $b_1$ fixed as $0.8$ while that $a_1$ fixed as $0.7$ in Fig.
6(b). In particular, as shown in Fig. 6, the proposed \emph{MRC
decoding scheme} outperforms the \emph{single signal decoding
scheme} for all the different cases. It is also illustrated that,
there exists a corresponding value of $a_1$ that optimizes the
performance in terms of the ergodic SR for both of two decoding
schemes. In addition, with the increase of the SNR, to achieve the
maximum ergodic SR, the power allocation factor $a_1$ will be
close to $1$.

\begin{figure}
\begin{center}
\includegraphics [width=110mm]{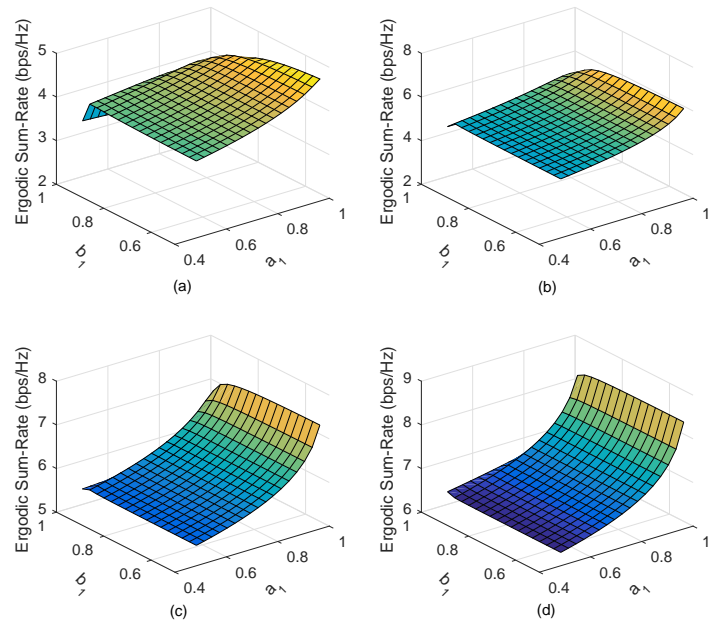}\\
\caption{\label{9} The ergodic SRs achieved by our proposed
\emph{Single Decoding scheme} with respect to the power allocation
factors $a_1$ and $b_1$, for (a): $\rho=15~\mathrm{dB}$, (b):
$\rho=20~\mathrm{dB}$, (c): $\rho=25~\mathrm{dB}$, (d):
$\rho=30~\mathrm{dB}$.}
\end{center}
\end{figure}

\begin{figure}
\begin{center}
\includegraphics [width=110mm]{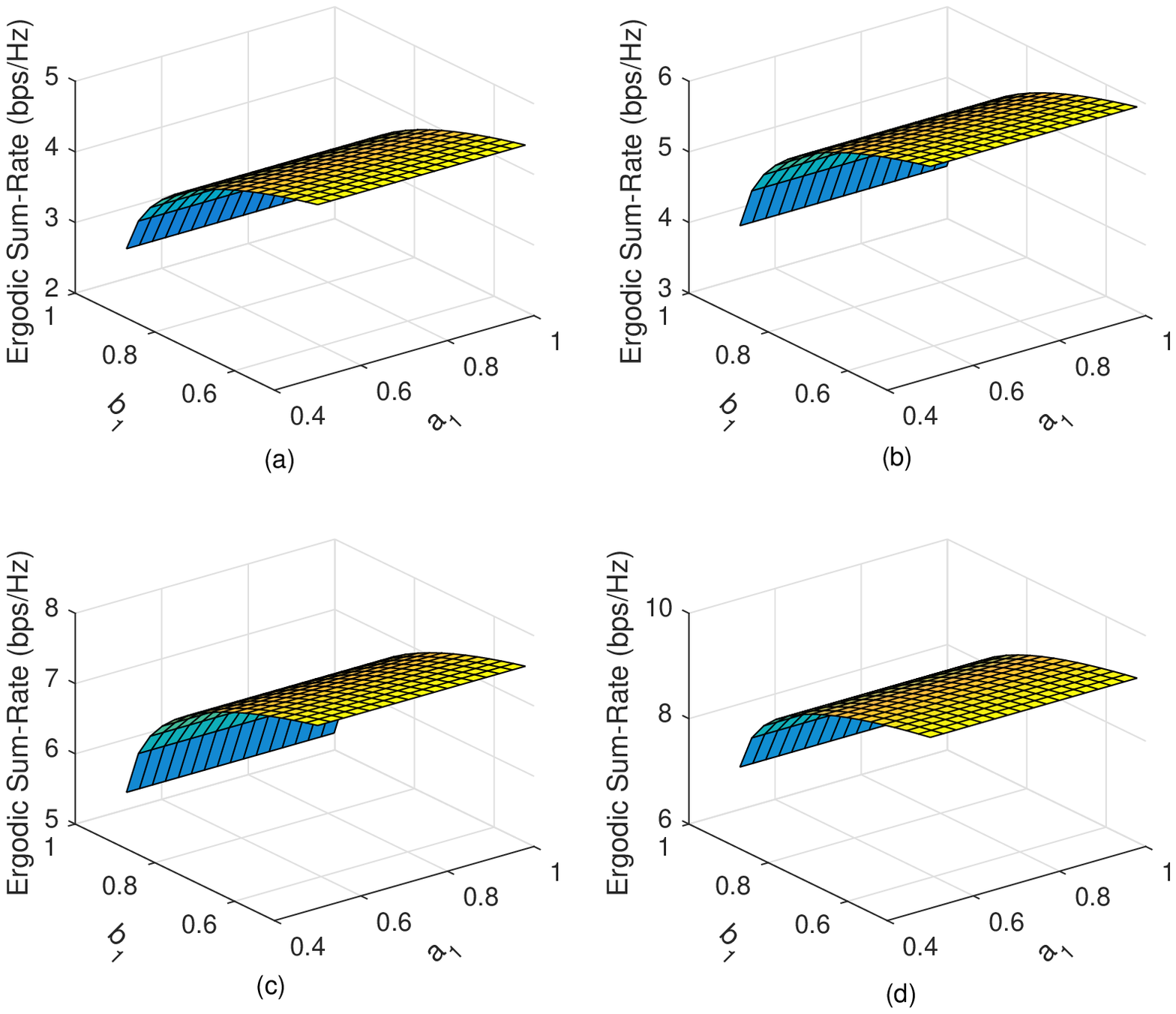}\\
\caption{\label{9} The ergodic SRs achieved by our proposed
\emph{MRC Decoding scheme} with respect to the power allocation
factors $a_1$ and $b_1$, for (a): $\rho=15~\mathrm{dB}$, (b):
$\rho=20~\mathrm{dB}$, (c): $\rho=25~\mathrm{dB}$, (d):
$\rho=30~\mathrm{dB}$.}
\end{center}
\end{figure}

Figs. 7-8 present the ergodic SR performance of our proposed
\emph{single signal decoding scheme} and \emph{MRC decoding
scheme} versus power allocation factors $a_1$ and $b_1$, where the
channel coefficients are set to be $\alpha_{SU_1}=5$,
$\alpha_{SU_2}=1$, $\alpha_{SU_3}=1$, $\alpha_{RU_2}=10$,
$\alpha_{RU_3}=10$ for different transmit SNR: $\rho=\left\{15,
20, 25, 30\right\}\mathrm{dB}$, respectively. Clearly, for both
figures, the maximum ergodic SR will be achieved for a value of
$a_1$ closed to ``1" while $b_1$ closed to ``0.5". To sum up, the
ergodic SR of the proposed \emph{MRC decoding scheme} provides an
outstanding advantage over the \emph{single signal decoding
scheme} and conventional D2D-NOMA one.

\section{Conclusions}

In this paper, a cooperative D2D-NOMA system has been studied with
two decoding strategies named the \emph{single signal decoding
scheme} and \emph{MRC decoding scheme}, respectively. The
asymptotic closed-form expressions for the ergodic SR, outage
probability and outage capacity for the proposed D2D-NOMA system
were provided. Specifically, numerical results have been presented
to corroborate the theoretical analysis, and results have
demonstrated that the \emph{MRC decoding scheme} yield significant
performance gains over the \emph{single signal decoding scheme}
and conventional D2D-NOMA one. Moreover, all the results showed
that the system performance are limited by the weak channel
condition for both the \emph{single signal decoding scheme} and
conventional NOMA scheme but not for the \emph{MRC decoding
scheme}. It remains future work to investigate the practical power
allocation method for the BS and users to further improve the system performance.\\\\
\textbf{List of Abbreviation
}\\
Cooperative relay networks (CRN), non-orthogonal multiple access
(NOMA), orthogonal multiple access (OMA), successive interference
cancellation (SIC), maximum ratio combining (MRC), sum-rate (SR),
power allocation (PA), frequency-division multiple access (FDMA),
time-division multiple access (TDMA), channel state information
(CSI), simultaneous wireless information and power transfer
(SWIPT), amplify-and-forward (AF), orthogonal frequency division
multiplexing (OFDM), coordinated direct and relay
transmission (CDRT), decode-and-forward (DF), additive white Gaussian noise (AWGN), base station (BS), internet of things (IoT), multiple-input multiple-output (MIMO). \\\\
\textbf{Availability of data and materials
}\\
The authors declare that all the data and materials in this manuscript are available.\\\\
\textbf{Competing interests}\\
The authors declare that they have no competing interests.\\\\
\textbf{Acknowledgements}\\ The author would like to thank the
Editor and reviewers
for their valuable and insightful comments.\\\\
\textbf{Authors' contributions}\\ W. Duan, J. Ju and G. Zhang
conceived and designed the study. W. Duan, Y. Ji, and Q. Sun
performed the simulations. W. Duan and Z. Wang wrote the paper. J.
Ju, G. Zhang, Z. Wang, Y. Ji and Q. Sun reviewed and edited the
manuscript. All authors read and approved the manuscript.

\end{document}